\documentclass{llncs}

\usepackage{amsmath,amsfonts,times,bbm,amscd,latexsym,amssymb}

\usepackage{amssymb}
\usepackage[mathscr]{euscript}

\def \ge		{\geqslant}      
\def \le		{\leqslant}      
\renewcommand\preceq {\preccurlyeq}
\renewcommand\succeq {\succcurlyeq}
\def \x			{\times}         
\def \N			{\mathbb N}      
\def \Z			{\mathbb Z}      

\def \implies			{\Longrightarrow}					
\def \union				{\,\cup\,}       
\def \inter				{\,\cap\,}       

\def \vide		{\varnothing}   
\def \et		{\,\wedge\,}    

\def \to		{\longrightarrow}                                 
\def \smallto		{\rightarrow}					  
\def \mapsto		{\longmapsto}					  
\def \subset		{\subseteq}					  
\def \epsilon		{\varepsilon}					  

\renewcommand\brace[1]	{\left\{ #1 \right\}}		       
\newcommand  \set[2]	{\brace{#1 \;\left|\; #2 \right.}}      
\newcommand  \abs[1]	{\left| #1 \right|}		       
\newcommand  \sqparen[1]{\left[ #1 \right]}		       

\newenvironment{romanum}
{\vspace{-1.5ex}
\let\jrnbtemptopsep\topsep
\setlength{\topsep}{-1ex}
\begin{enumerate}
	\leftskip=1em
	\renewcommand\theenumi{\textup{(\textit{\roman{enumi}})}}
	\setlength{\itemsep}{-0.2ex}
}{\end{enumerate}\vspace{-1ex}
\setlength{\topsep}{\jrnbtemptopsep}}

\newcommand\arc\smallto

\newcommand\mC{\mathcal{C}}
\newcommand\mF{\mathcal{F}}

\newcommand\mA{\mathcal{A}}
\newcommand\mB{\mathcal{B}}

\newcommand\flowi{\textup{(\textsf{F}\textit{\!\!\;i})}}
\newcommand\flowii{\textup{(\textsf{F}\textit{\!\!\;i\!\!\;i})}}
\newcommand\flowiii{\textup{(\textsf{F}\textit{\!\!\;i\!\!\;i\!\!\;i})}}

\newcommand\ri{\textup{\textbf{(i)}}}
\newcommand\rii{\textup{\textbf{(ii)}}}

\newcommand\typ[1]{\textup{\textbf{\texttt{#1}}}}

\def\qed{$\Box$}

\def\PF{\noindent \textbf{Proof.}~~}
\def\FP{\hfill \qed \\}

\def\card#1{|#1|}

\def\ens#1{\{#1\}}
\def\ie{\textit{i.e.}}
\def\eg{\textit{e.g.}}

\def\mathvec#1{\mathbf{#1}}
\def\al{\alpha}

\def\ta{\theta}

\def\rar{\rightarrow}

\def\mcl{\mathcal} 
\def\mbb{\mathbb}

\newtheorem{prop}{Proposition}\def\PRO{\begin{prop}}\def\ORP{\end{prop}}
\newtheorem{coro}{Corollary}\def\COR{\begin{coro}}\def\ROC{\end{coro}}
\newtheorem{theo}{Theorem}\def\TH{\begin{theo}}\def\HT{\end{theo}}
\newtheorem{defi}[prop]{Definition}\def\DE{\begin{defi}}\def\ED{\end{defi}}
\newtheorem{lemme}[prop]{Lemma}\def\LE{\begin{lemme}}\def\EL{\end{lemme}}
\newtheorem{algo}{Algorithm}
\def\AL{\begin{algo}}\def\LA{\end{algo}}
\newcommand{\AR}[2][c]{$$\begin{array}[#1]{lllllllllllllll}#2\end{array}$$}
\def\MA#1{\left(\begin{matrix}#1\end{matrix}\right)}
\def\EQ#1{\begin{eqnarray}#1\end{eqnarray}}

\def\ket#1{{|}#1\rangle}
\def\bra#1{\langle#1{|}}
\def\ctR{\mathop{\wedge}\hskip-.4ex}
\def\ctwo{{\mbb C}^2}

\def\ost{\frac1{\sqrt2}}

\def\pit{\frac\pi2}
\def\Cx#1{\cx{#1}{}}
\def\Cz#1{\cz{#1}{}}
\def\cz#1#2{Z_{#1}^{#2}}
\def\cx#1#2{X_{#1}^{#2}}
\def\ei#1{e^{i#1}}
\def\emi#1{e^{{-i}#1}}
\def\cx#1#2{X_{#1}^{#2}}
\def\phZ#1#2{N_{#1}^{#2}}

\def\m#1#2{{M}_{#2}^{#1}}
\def\M#1#2{{M}_{#2}^{#1}}
\def\et#1#2{E_{#1#2}}

\def\del{\,\partial\,}

\begin{document}

\title{Phase map decompositions for unitaries}  
\author{Niel de Beaudrap\inst{1}, Vincent Danos\inst{2}, and Elham Kashefi\inst{1}}
\institute{IQC, University of Waterloo \and CNRS, Universit{\'e} Paris 7}

\maketitle

\begin{abstract}
We propose a universal decomposition of unitary maps over a tensorial power of $\ctwo$, introducing the key concept of \emph{phase maps}, and investigate how this decomposition can be used to implement unitary maps directly in the measurement-based model for quantum computing. 
Specifically, we show how to extract from such a decomposition a matching entangled graph state (with inputs), and a set of measurements angles,
when there is one. Next, we check whether the obtained graph state verifies a \emph{flow} condition, which guarantees an execution order such that the dependent measurements and corrections of the pattern yield deterministic results. Using a graph theoretic characterization of flows, we can determine whether a flow can be constructed for a graph state in polynomial time. This approach yields an algorithmic procedure which, when it succeeds, may produce an efficient pattern for a given unitary.
\end{abstract}

\section{Introduction}

Measurement-based quantum computing \cite{Gott99,RB01,mqqcs,Nielsen03} has attracted considerable interest in quantum computing research during the last few years. Different aspects of this model
have been investigated, including: characterization of the initial entangled states \cite{Schl03,graphstates,DAB03,VDD04b,VDD04,MP04}, description and unification of different
models \cite{L03,AL04,CLN04,PJ04,JP04,DKP04,dk05a,RV05}, investigation of fault tolerant computing within this framework \cite{RAB04,ND04,RBH04,PL05,VBR05,RHG05,DHN05}, and proposals for implementations \cite{Nielsen04,CMJ04,BR04,TPKV04,TPKV05,nature05,KPA05,BES05,CCWD05,BBFM05}.
 
 It remains an open question however, whether this model may suggest new techniques for designing quantum algorithms. This is the question we address in the present paper. Specifically, we introduce a methodology for the direct decomposition of a given unitary map into a one-way model pattern.

Previously, one would typically start with an algorithm already implemented in the circuit model,
replace each gate by a corresponding pattern. To transform the pattern to a standard form where all entangling operations are performed first, one could then use the by-product method~\cite{mqqcs} or the more general standardization algorithm~\cite{DKP04}. In this paper, we propose a direct method that is free from any reference to the circuit model. We start with the observation that one-way patterns implicitly define a particular decomposition of unitary maps into a preparation map enlarging the input space, a diagonal map with unit coefficients, and a restriction map contracting back the space to the output space, which we call a \emph{phase map decomposition}. (Note that this decomposition does not directly correspond to any physical procedure.)

Since the one-way model is universal, this alternative decomposition is also universal; and there is a straightforward procedure which allows us to determine a phase map decomposition for a unitary from a pattern implementing the same unitary. Remarkably, one can define a reverse procedure as well, which breaks in two steps. First, given a unitary map, one enumerates such phase map decompositions by constructing the right set of coefficients in the middle diagonal map (Lemma~\ref{L-eq} and Algorithm \ref{A-phase}). Then for each such decomposition, one verifies whether there exists a matching one-way pattern. This reduces to finding for any phase map decomposition a matching entangled graph state (with inputs) and choice of measurements angles (Algorithm \ref{A-graph}). To obtain the execution order of the one-way pattern, we look for a ``\emph{flow}'' for the underlying entanglement graph which describes an adequate set of dependent measurements and corrections so that the obtained pattern is deterministic (Theorem~\ref{t-flow}). Using a graph theoretic characterization of flows, we can determine whether a flow can be constructed for a graph state in polynomial time (Algorithms~\ref{alg:findCover} and~\ref{alg:findDepthFn}).

This constitutes a method for the implementation of unitary maps in the one-way model which
bypasses completely the circuit picture, and may produce more efficient implementations in particular instances.

\section{Preliminaries}

We first briefly review the one-way model and the various notions relevant to  this paper.

\subsection{One-way Model}

Computations involve a combination of one-qubit preparations $N_i$, two-qubit entanglement operators $\et ij:=\ctR Z_{ij}$,\footnote{The symmetric operator of controlled-$Z$.} one-qubit measurements $\M\al i$, and one-qubit Pauli
corrections $\Cx i$, $\Cz i$ (where $i$, $j$ represent the qubits on which each of these operations apply, and $\al$ is a parameter in $[0,2\pi]$). A preparation $N_i$ prepares auxiliary qubit $i$ in state $\ket{+}_i$. An entangling operator $\et ij$ entangles qubits $i$ and $j$. A destructive measurement $\M\al i$ is defined by the pair of orthogonal projections $\bra{\pm_\al}_i$ applied at qubit $i$, where $\ket{\pm_{\al}}$ stand for $\ost(\ket0\pm\ei{\al}\ket1)$. We use the convention that $\bra{+_{\al}}_i$ corresponds to the outcome $0$, and $\bra{-_{\al}}_i$ corresponds to $1$. Since qubits are measured at most once in a pattern, we may unambiguously represent the outcome of the measurement done at qubit $j$ by $s_j$.
The dependent corrections used to control non-determinism are then $\cx i{s_j}$ and $\cz i{s_j}$, with $s_j \in \{0,1\}$.

A \emph{pattern} is defined by the choice of a finite set $V$ of qubits, two possibly overlapping subsets $I$ and $O$ determining the pattern inputs and outputs, and a finite
sequence of commands acting on $V$. We require that  no command depends on an outcome not yet measured; that no command acts on a qubit already measured; and that a qubit is measured (or prepared) if and only if it is not an output (respectively an input). 

Let $\mcl H_I$ ($\mcl H_O$) denote the Hilbert space spanned by the inputs (outputs). The execution of a pattern consists of performing each command in sequence. If $n$ is the number of measurements (\ie\ the number of non-output qubits), then this may follow $2^n$ different computational branches. Each branch is associated with a unique binary string  $\mathvec s$ of length $n$, representing the classical outcomes of the measurements along that branch, and a unique \emph{branch map} $A_{\mathvec s}$ representing the linear transformation from $\mcl H_I$ to $\mcl H_O$ along that branch.  (Note that states resulting from a measurement \emph{are not} normalized.) The branch corresponding to the classical outcomes $\mathvec s=0\cdots0$, will be called the \emph{positive branch}.

A pattern is said to be \emph{deterministic} if all of the branch maps are equal: any given unitary can be implemented by a deterministic pattern (in the precise sense of Theorem~\ref{t-flow} below). A pattern is said to be in \emph{standard form} if entangling operators $\et ij$ appear first in its command sequence. Any given pattern can be put in such a form, which reveals implicit parallelism in the computation, and suits certain implementations well~\cite{DKP04}. One can then define an \emph{entanglement graph} associated to the initial preparation and entanglement step, with vertices $V$ and edges $\et ij$.  By doing so, the computation of the pattern can be understood as the direct manipulation of such multipartite entangled states (which are called graph states in the case where there are no inputs). 

\subsection{Geometries and Flows}
In order to find a pattern for a given phase map decomposition of a unitary operator, we will need to search for a matching entanglement graph and write the pattern directly from the obtained graph. We may do this using \emph{flows}~\cite{dk05c}, which we review now.

A \emph{geometry} $(G,I,O)$ consists of an undirected graph $G$ together with two subsets of nodes $I$ and $O$, called inputs and outputs. We write $V$ for the set of nodes in $G$, $I^c$, and $O^c$ for the complements of $I$ and $O$ in $V$\,. We write $x \sim y$ and $xy \in G$ when $x$ is adjacent to $y$ in $G$\,, and $E_G := \prod_{xy \in G}E_{xy}$ for the global entanglement operator associated to $G$. (Note that all $E_{xy}$ commute so the order in this product is irrelevant.) To extend a geometry into a pattern, it remains only to decide the angles which are used in measuring qubits of $O^c$, and the dependent corrections to be used; thus every pattern has an underlying geometry which is obtained by forgetting the measurements and corrections.

We give now a condition on geometries under which one can synthesize a set of dependent corrections such that the obtained pattern is deterministic.

\DE
A \emph{flow} $(f,\preceq)$ for a geometry $(G,I,O)$ consists of a map $f:O^c\rar I^c$ and a
partial order $\preceq$ over $V$ such that for all $x\in O^c$:
\\---~~\flowi~~$x \sim f(x)$;
\\---~~\flowii~~$x \preceq f(x)$;
\\---~~\flowiii~~for all $y \sim f(x)$\,, we have $x \preceq y$\,.
\ED
The coarsest order $\preceq$ for which $(f, \preceq)$ is a flow will be called the
\emph{dependency order} induced by $f$.

\TH[\cite{dk05c}] \label{t-flow}
Suppose the geometry $(G,I,O)$ has flow $f$, then the pattern:
\AR{
\mcl P_{f,G,\vec\al}&:=&\prod\limits_{i\in O^c}{\!\!\!}^{\preceq}\,\,\,
\Big(\cx{f(i)}{s_i}\prod\limits_{\substack{k \sim f(i) \\ k \ne i}}\cz{k}{s_i}\m{\al_i}i\Big)
E_G \phZ {I^c}{}
}
where the product follows the dependency order $\preceq $ of $f$,  is deterministic, and realizes the unitary embedding:
\AR
{U_{G,I,O,\vec\al}&:=&
{2}^{|O^c|/2}\;\Big(\prod\limits_{i\in O^c}\bra{{+_{\al_i}}}_i \;\Big)\, E_G\phZ {I^c}{}}
\HT

That is: if a geometry has a flow, we may without loss of generality consider only its positive branch. Then measurement-based computing downs to \emph{projection-based} computing. This will be handy in the formulation of our procedure.

\section{One-way Patterns as Phase Maps}
With our preliminaries in place, we now turn to the formulation of our decomposition. Various operators over $\mcl H$ preserve the computational basis, up to phase: for example, Pauli maps, $Z^\al$ (defined as $Z^\al\ket0=\ket0$, and $Z^\al\ket1=\ei\al\ket1$), and controlled-Paulis. One-qubit measurements also map the standard basis of one space to those of another, up to a scalar factor. In particular, one has the following simple equations where $\ket{x}$ is an $n$-qubit computational basis state and $j$ is an index for a qubit in $\ket x$:
\AR{
\bra{\pm_\alpha}\otimes I^{\otimes n}(\ket0\ket x)&=&2^{-1/2}\ket x\\
\bra{\pm_\alpha}\otimes I^{\otimes n}(\ket1\ket x)&=&\pm2^{-1/2}e^{-i\al}\ket x\\
\ctR Z_{1j}\ket0\ket x&=&\ket0\ket x\\
\ctR Z_{1j}\ket1\ket x&=&Z_j\ket1\ket x\\
}
We will call a map $\Phi:\mcl H_V\to\mcl H_V$ a \emph{phase map} if it is diagonal in the computational basis and has only unit coefficients. The typical example of such a map is $\ctR Z$. It is important to note that the above definition depends on the choice of a basis.

Following one-way model terminology we also define a \emph{preparation map} $P_{I\rightarrow V} : \mcl H_I \rightarrow \mcl H_V$ that expands the input space by tensoring auxiliary qubits,
\AR{
\ket x \mapsto \ket x \otimes \ket{+\cdots+}_{I^c}	\;\;,
}
and a \emph{restriction map} $R_{V \rightarrow O} : \mcl H_V \rightarrow \mcl H_O$ that projects the space to the output space:
\AR{
\ket x \mapsto \bra{+\cdots+}_{O^c} \ket x	\;\;.
}
It is easy to see that the restriction map is the adjoint of the preparation map.

As we have seen above, measurement and entangling commands in the one-way model define phase maps, and hence from the universality of the model we obtain the following decomposition:
\TH\label{t-decompo-app}
For all unitary $U:\mcl H_I\to\mcl H_O$, there exists a phase map
$\Phi:\mcl H_V\to\mcl H_{V}$ such that:
\AR{ 
U=R_{V\rightarrow O}\circ \Phi\circ P_{I\rightarrow V}
}
\HT
\PF We know that $\ctR Z$ and $J_\al$ are universal~\cite{generator04}, and also that compositions of patterns having flows, themselves have a flow~\cite{dk05c}. Hence there is a deterministic pattern $\mcl P$ implementing $U$, and from Theorem \ref{t-flow} we then have:
\AR{
U&=&
2^{\card{O^c}/2}\;\prod_{i\not \in O} \bra{+_{\al_i}}_i \;E_G\; P_{I\rightarrow V}\\ \\
&=&2^{\card{O^c}/2}\;\prod_{i\not \in O} \bra{+}_i Z^{-\al_i}_i \; E_G \; P_{I\rightarrow V}\\ \\
&=&R_{V\rightarrow O}\;\prod_{i\not \in O} Z^{-\al_i}_i E_G \; P_{I\rightarrow V}\\ \\
&=&R_{V\rightarrow O}\;  \Phi_{V\rightarrow V} \; P_{I\rightarrow V}
}
where $\Phi = \prod_{i\not \in O} Z^{-\al_i}_i \prod_{ij \in E} \ctR Z ij $ is the phase map corresponding to the entanglement operations and measurement angles.
\FP

One can think of the above theorem as a special kind of diagonalization for unitaries where one is allowed to inflate the dimension of the underlying space. This will prove to be useful for direct programming in the one-way model. We present first a couple of examples, clarifying our first set of definitions and preparing the ground for a direct proof of the above theorem which does not invoke universality of the one-way model. This then leads to an algorithm for pattern design. The examples already hint at the construction behind the direct phase map decomposition algorithm.

\subsection{Example 1}
Consider the unitary map $J_\al:\mcl H_{\{1\}}\to \mcl H_{\{1\}}$ which decomposes in the
computational basis as:
$$
J_\al
=
2^{-1/2}
\MA{
1&e^{-i\al}\\
1&-e^{-i\al}
}
=
\MA{
\;1\;&\;0\;&\;e^{-i\al}&\;0\;\\
\;0\;&\;1\;&\;0\;&-e^{-i\al}
}
\cdot
2^{-1/2}
\MA{
1&0\\
1&0\\
0&1\\
0&1
}
$$
This decomposition is obtained from the one-way pattern $\cx 2 {s_1}\m \al 1\et 12$ which implements $J_\al$, and has as positive branch the phase map:  
\AR{
2^{1/2}\bra{+_\al}_1\ctR Z_{12}:\mcl H_{\ens{1,2}}\to\mcl H_{\ens{2}}
} 
Factoring out the restriction operator gives the decomposition:
\AR{
J_\al&=&
\MA{
1&0&1&0\\0&1&0&1
}
\cdot
\MA{
\;1\;&\;0\;&\;0\;&\;0\;\\
\;0\;&\;1\;&\;0\;&\;0\;\\
\;0\;&\;0\;&e^{-i\al}&\;0\;\\
\;0\;&\;0\;&\;0\;&-e^{-i\al}
}
\cdot
2^{-1/2}
\MA{
1&0\\
1&0\\
0&1\\
0&1
}
}
where the left matrix is the restriction $R_{1}:\mcl H_{\ens{1,2}}\to\mcl H_{\ens{2}}$.

Note that this decomposition is not physical, since the phase map doesn't directly correspond to any operation. Indeed it is a convention to substitute the physically meaningful equation
$\bra{+_\al}\otimes I( \ket 1\ket 1)=2^{-1/2}e^{-i\al}\ket 1$ with:
\AR{ 
R_{1}(2^{-1/2}e^{-i\al}\ket 1\ket 1)=2^{-1/2}e^{-i\al}\ket 1
}
However, the corresponding one-way pattern is a physical procedure that may be understood in terms of the above decomposition.

The phase map here is $Z_1^{-\al}\ctR Z_{12}$, and the decomposition above can be rewritten:
\AR{
J_\al
&=&R_1(Z_1^{-\al}\ctR Z_{12})P_2\\
}
As we have seen in the proof above, this reasoning is perfectly general; indeed 
for any pattern, one gets the phase map associated to the positive branch:
\AR{ \Phi&=&\prod\limits_{i\in O^c}Z_i^{-\al_i}\prod\limits_{ij\in E}\ctR Z_{ij}
}
and writing $U$ for the corresponding unitary (assuming the pattern does compute a unitary map):
\AR{
U
&=&
R_{O^c}
\,\Big(\prod\limits_{i\in O^c}Z_i^{-\al_i}\prod\limits_{ij\in E}\ctR Z_{ij}\,\Big)\,
P_{I^c}
}

\subsection{Example 2}
Example 1 uses only one auxiliary qubit, and as such is a special case where the required number of auxiliary qubits is equal to the number of inputs. This is of course not always the case and the general algorithm for phase map decomposition will take care of this. We will present exact bounds on how much one needs to expand the computational space to be able to obtain the decomposition; however to realize a decomposition as a pattern we will need further restrictions. The following example demonstrate this case. 
 
The shortest known pattern for the $Z^{\al}$-rotation is $\cx 3{s_2} \cz 3{s_1} \m 02 \m {-\al}1\et 12\et 23$ with positive branch:
\AR{
\bra{+}_2\bra{+_{-\al}}_1\et12\et23
}
which induces the 3-qubit phase map $\Phi \ket{xyz}={(-1)}^{xy+yz}\ei{\al y}\ket{xyz}$ (indeed a diagonal of units) and corresponds to the following decomposition of $Z^\al$:
\AR{
R_{12}\,D(1,1,1,-1,\ei\al,\ei\al,-\ei\al,\ei\al)\,P_{23}
}
where $D(\cdots)$ is a diagonal matrix.  Note that some permutations of the diagonal lead to other solutions, and most decompositions won't correspond to a pattern. Also if one uses only 1 additional qubit one obtains another decomposition with middle map:
\AR{D(\sqrt2,0,0,\sqrt2\ei\al)}
which is not a phase map since coefficients are not units. The natural question is whether it is possible to generate all such decompositions without using a pattern, and next find a pattern matching one of these decompositions. 

\section{Direct Decomposition}

There is no \emph{a priori} reason why a phase map decomposition could not be obtained without any reference to the one-way model: indeed, it is possible to give a direct approach. Supposing one adds $n$ auxiliary qubits to the input space $I$, a simple calculation shows that for each coefficient $u$ in the the computational basis matrix representation of $U$, there will be ${2^{\card V - \card O}}/{2^{\card I}}=2^{n-\card I}$ `slots' to spread over the diagonal of the phase map $\Phi$ (since $U$ is a unitary $\card I = \card O$). Thus, finding a decomposition amounts in this case to finding complex numbers $x^{(i)}$
such that the following two conditions hold:
\EQ{
u=\sum_{i\leq2^{n-\card I}}x^{(i)}\label{e-add}
\\
2^{n/2}|x^{(i)}|=1 \label{e-norm}
}
The first equation says that the restriction map $R$ will sum up all the $x^{(i)}$s to give $u$, while the second one asks for unit diagonal elements (note that the preparation map of $n$ auxiliary qubits introduces an overall factor of $2^{-n/2}$).
\LE \label{L-eq}
If $n > |I|$\,, Equations~\ref{e-add} and~\ref{e-norm} have joint solutions iff $|u| \le 2^{n/2-\card I}$\,.
\EL
\PF Each complex $x^{(i)}$ can be seen as a real plane vector of constant length ${{2^{-n/2}}}$, and all one has to do is to choose their angles in a way that they will globally add up to $u$. If one aligns all $x^{(i)}$s with $u$, the resulting sum is at least as long as $u$ iff $|u| \le 2^{n/2 - |I|}$\,; thus this inequality is necessary for Equations~\ref{e-add} and~\ref{e-norm} to have joint solutions.
 
If $n > |I|$\,, then $2^{n - |I|} \ge 2$\,, so there are at least two terms $x^{(i)}$\,. We may pick any two of them and rotate them at opposite angles $\pm\ta$. If $\ta$ reaches $\pit$ before the global sum matches $u$, then the corresponding two $x^{(i)}$s contribute nothing, and we pick two additional terms to rotate. Clearly, at some stage, for some value of $\ta$ the sum will coincide with that of $u$.
\FP

Due to the unitarity of $U$, $|u|\leq 1$\,: so a safe choice of  $n$ is one such that $2^{n/2-\card I} \geq 1$. Thus, $n\geq2 \card I$ is always sufficient for a phase map to exist. Another consequence of the above lemma is that for any given unitary map $U$ on $|I|$ qubits, unless $U$ is itself a phase map and also requires no auxiliary qubits, we have $n > |I|$: then a lower bound on the number of required qubits to implement it as a one-way pattern is $2|I|$, if at least one coefficient of $U$ is larger than $\frac{1}{2}$.

For a unitary $U$\,, once we have fixed $n$, an output space $O$, and a choice of $x^{(i)}_{pq}$ satisfying Equations~\ref{e-add} and~\ref{e-norm} for the coefficients $u_{pq} = \bra p U \ket q$, the following algorithm will enumerate all possible decompositions:
\AL \label{A-phase}
\emph{Input}: for sets $V$, $I$, and $O$ and $n=\card {I^c}$\,:
\\---~~a unitary $U$ on $\mcl H_I$;
\\---~~complex numbers $\{\;x^{(i)}_{pq}\;\}_{i = 1}^{2^{n - \card I}}$ satisfying Equations~\ref{e-add} and~\ref{e-norm} for each $u_{pq}$\,;
\\---~~a permutation $\sigma$ over $\{1, \cdots, 2^{n-\card I} \}$.\\

\emph{Output}: diagonal elements $\{ d_{kk} \}_{k = 1}^{2^{|V|}}$\,, such that
$d_{kk} = \sqrt{2^n}\,x^{(i)}_{pq}$\,, where:
\\---~~the binary representation of $p$ agrees with that of $k$ after restriction to $O$;
\\---~~$q \equiv k \;{\rm mod}\; 2^{\card I}$;
\\---~~$i = \sigma(\lfloor k/2^{\card I}\rfloor)$.
\LA

The elements $\{ d_{kk} \}_{k = 1}^{2^n}$ are the solution of $R \Phi P \,=\, U$
where \;$\Phi = D(d_{11}, d_{22}, \cdots)$\,; due to the simple structure of matrices $R$
and $P$ we derive the above algorithm.

Note that obtaining a decomposition is not sufficient for the existence of a pattern. We now discuss how to extract an entanglement graph and measurement angles from a phase map, if there is one.

\section{Pattern Design}
To determine whether a phase map decomposition $R \,\Phi\, P$ of a unitary $U$ has a corresponding pattern in the one-way measurement model, one wants a graph $G_E$ over 
$V$, and angles $\al_j$ for $j\in O^c$ such that 
\AR{\Phi=\prod\limits_{j\in O^c}Z_j^{-\al_j}\prod\limits_{jk\in E}\ctR Z_{jk}} 
That means for all $x$ in the $V$-computational basis:
\EQ{
\label{e-patt}
d_{xx}=\emi{\sum_{O^c}\al_jx_j}\,(-1)^{\sum_{jk\in E}x_jx_k}
}
where $d_{xx}$ is the diagonal coefficient of the phase map corresponding to $x$ basis. Based on this observation we propose the following algorithm for the above graph matching problem.

\AL \label{A-graph}
\emph{Input}: A phase map decomposition for $U$ ---
\ie\ the diagonal elements $\{ d_{xx} \}_{x = 1}^{2^{|V|}}$ from Algorithm~\ref{A-phase}.
\\\emph{Output}: either \ri~A~graph $G$ on $V$ and a set of angles of measurements
$\{\al_j\}_{j \in O^c}$, or \rii~no~matching graph exists.
\begin{enumerate}
\item
	For \;$j \in \{ 1, \cdots, \card{O^c} \}$, consider the $\card{V}$-bit string
	$\mathbf{z}_j$ that only has a $1$ at position $j$, and set $\al$ such that
	$\emi{\al_j}=d_{\mathbf{z}_j \mathbf{z}_j}$\,.
	
\item
	For all $j,k$, consider the $\card{V}$-bit string $\mathbf{z}_{jk}$ having a $1$ only at
	positions $j$ and $k$. Check whether $d_{\mathbf{z}_{jk} \mathbf{z}_{jk}} =
	 \pm\emi{(\al_j+\al_k)}$ (the angles for the corresponding qubit in $O$ is taken to be 0).
	\\---~~\ri~if~YES and the sign is $-1$, return $\et jk$ as an edge in $G$.
	\\---~~\rii~if~NO, no matching graph exists.
\end{enumerate}
\LA

From Equation~\ref{e-patt}, we can see that a given phase map can be implemented by at most one pattern. Once an entanglement graph is obtained, we check whether $(G,I,O)$ has a flow, in which case Theorem \ref{t-flow} obtains a deterministic pattern for $U$. We address the problem of finding a flow for the entanglement graph in the next section.

Although there are exponentially many elements on the diagonal of the phase map, testing for the existence of $G$ will query the middle diagonal map only quadratically in $|I|+n$, to read off the measurement angles and the entanglement graph. This in practice could accelerate the detection of bad decompositions before obtaining all diagonal elements in the phase map.

When the procedure fails, because the obtained graph does not have flow or simply does not exist, one backtracks by: \textbf{(1)}~trying a different decomposition given by Algorithm~\ref{A-phase}, \textbf{(2)}~trying another solution from Lemma~\ref{L-eq}, \textbf{(3)}~revising the choice of outputs, and ultimately \textbf{(4)}~expanding further the computational space. Without any additional constraints, it seems that there are many solutions to be checked. One might be able to infer additional constraints to Equations~\ref{e-add} and~\ref{e-norm} from the requirement that there be a corresponding entanglement graph, reducing the set of phase maps which we consider. How this may be done is, however, an open question.

\section{Finding Flows}

In this section, we outline algorithms which allow us to find flows for a geometry $(G,I,O)$
in polynomial time. The proof of correctness of the algorithm employs technical results
of graph theory, which we will only sketch here: a rigorous development may be found
in~\cite{Beaudrap06}.

\subsection{Graph Theoretic Characterization of Flows}

We will begin by showing how flows can be characterized in the language of graph theory,
in order to make use of the solved problems in that field.

First, we present some additional notation related to directed paths. In a directed path,
we write $x \arc y$ to represent an arc between vertices $x$ and $y$. If $\mC$ is a collection
of directed paths, we will say that $x \arc y$ is an \emph{arc of $\mC$} when $x \arc y$ is
an arc in one of the elements of $\mC$\,.

For a flow $(f, \preceq)$\,, we can easily see that the function $f$ must be injective:
if $f(x) = f(y)$\,, we have $y \sim f(x)$ and $x \sim f(y)$ by \flowi, and $x \preceq y$
and $y \preceq x$ by \flowiii, so $x = y$\,. Therefore, each vertex $x \in V$ has at most
one vertex $w \in V$ such that $x = f(w)$ and hence the orbits of vertices under the
function $f$ induce a path-like structure.
\DE
	\label{dfn:pathCover}
	Let $(G,I,O)$ be a geometry. A collection $\mC$ of (possibly trivial) directed paths in
	$G$ is a \emph{path cover} of $(G,I,O)$ if
	\begin{romanum}
	\setlength\itemsep{0.5ex}
	\item
		each vertex $v \in V$ is contained in exactly one path;
	\item
		each path in $\mC$ is either disjoint from $I$\,, or intersects $I$ only at its
		start point;
	\item
		each path in $\mC$ intersects $O$ only at its end point.
	\end{romanum}
\ED

A path cover of $(G,I,O)$ is meant to capture the way that information flows within the
geometry during the execution of a pattern. When $\abs{I} = \abs{O}$, a path cover is just
a maximum-size collection of directed paths from $I$ to $O$ which do not intersect
(\ie\ a \emph{maximum family of vertex-disjoint $I$--\;\!$O$ paths}) which also happens
to cover every vertex of the graph.

Every flow function $f$ generates a path cover from its' orbits, and conversely we can obtain
a natural function $f$ from any path cover. However, not all path covers correspond to flows,
because the function $f$ from an \emph{arbitrary} path cover may not have a compatible
partial order satisfying \flowii\ and \flowiii. For example, consider the geometry given
by alternating input and output vertices on a cycle of length $6$:
\begin{itemize}
\setlength\itemsep{0.5ex}
\item
	$V = \brace{0,1,2,3,4,5}$\,, $I = \brace{0,2,4}$\,, and $O = \brace{1,3,5}$\,;
\item
	there is an edge between $i, j \in V$ iff $j-i \equiv \pm 1 \pmod{6}$\,.
\end{itemize}
The only two path covers for $(G,I,O)$ are generated by the functions
$f(x) \equiv x+1 \pmod{6}$ and $f(x) \equiv x-1 \pmod{6}$\,. In the former case,
we have $c_2 \sim f(c_0)$\,, $c_4 \sim f(c_2)$\,, and $c_0 \sim f(c_4)$\,: then,
from condition \flowiii, we would require $c_0 \preceq c_2 \preceq c_4 \preceq c_0$\,.
Similarly, in the latter case, we would have $c_0 \succeq c_2 \succeq c_4 \succeq c_0$\,.
However, neither of these are possible if $\preceq$ is a partial order.

In order to characterize when a path cover corresponds to a flow, we must introduce a
condition which prohibits such a cycle of relations between distinct vertices. We
do this by capturing the conditions which make a path cover give rise to these
relations in the first place.
\DE
	\label{dfn:viciousCircuit}
	Let $\mF$ be a family of directed paths in a graph $G$\,. A closed walk $C = c_0 c_1
	\cdots c_{m\!-\!1} c_0$ (indexed over $\Z_m$) in $G$ is a \emph{vicious circuit for $\mF$}
	if for all $j \in \Z_m$\,, $c_{j\!-\!1} \ne c_{j\!+\!1}$\,, and at least one of
	$c_{j\!-\!1} \arc c_j$ or $c_j \arc c_{j\!+\!1}$ is an arc of $\mF$\,.
\ED
That is: a vicious circuit is a walk that only visits vertices covered by $\mF$\,, never
doubles back on an edge which has just been traversed, and which traverses an arc of $\mF$
on at least one out of every two consecutive edges. The idea behind considering such a
structure is that such a vicious circuit can be cut into segments, each consisting of
one or two edges:
\begin{itemize}
\setlength\itemsep{0.5ex}
\item
	\textbf{One edge segment:} an edge $xy$\,, where $x \arc y$ is an arc of $\mF$\,;
\item
	\textbf{Two edge segment:} a path $xyz$\,, where $x \arc y$ is an arc of $\mF$\,,
	but $y \arc z$ is not.
\end{itemize}
The importance of these segments is as follows. Let $\mC$ be a path cover, and $f$ be the
injective function naturally coming from it: then $x \arc y$ is an arc of $\mC$ iff
$y = f(x)$\,. For any $x$ and $y$ which are the initial and final points of such a segment,
any binary relation $\preceq$ satisfying \flowii\ and \flowiii\ will have $x \preceq y$\,.
As vicious circuits can be decomposed into such segments, a vicious circuit will then
induce the cyclic relations we wish to prohibit. This motivates the following definition:
\DE
	\label{dfn:causalPathCover}
	A \emph{causal path cover} of $(G,I,O)$ is a path cover which does not have any
	vicious circuits in $G$\,.
\ED
The definition above is intended to simply exclude those path covers which induce
cyclic relations of the sort above. In fact, this definition characterizes flows:
\TH[\cite{Beaudrap06}, Corollary 11]
	\label{thm:graphth-charn}
	A geometry has a flow iff it has a causal path cover.
\HT

\subsection{Uniqueness of Causal Path Covers among Families of Disjoint $I$--\;\!$O$ paths}

Theorem~\ref{thm:graphth-charn} tells us that in order to find a flow function $f$
for a geometry $(G,I,O)$\,, we may replace the condition that $f$ has a compatible
partial order satisfying \flowii\ and \flowiii\ with the condition that the path
cover induced by $f$ lacks vicious circuits. In the special case where $\abs{I} = \abs{O}$
(\eg\ for a geometry from the pattern for a unitary transformation), we have the following
result:
\TH[\cite{Beaudrap06}, Theorem 16]
	\label{thm:causalUniquePathFamily}
	Let $(G,I,O)$ be a geometry such that $\abs{I} = \abs{O}$\,. If $(G,I,O)$ has a causal
	path cover $\mC$\,, then it is the only maximum family of vertex-disjoint $I$--\!\;$O$
	paths in $G$\,.
\HT
The reason for this is that any other maximum family of vertex-disjoint $I$--\;\!$O$ paths
either fails to cover all vertices of $G$, or traverses a different set of edges in $G$\,.
From this, one can prove that a vicious circuit must exist for $\mC$\,. Thus, if $\mC$ is
causal, no other maximum family of vertex-disjoint $I$--\;\!$O$ paths may exist. As a
corollary, because a causal path cover for $(G,I,O)$ is unique when it exists, a geometry
$(G,I,O)$ has at most one possible flow function.
 
Using the above result, if $(G,I,O)$ has a flow, we may find it by first finding any maximum
family of vertex-disjoint $I$--\;\!$O$ paths, which has a known reduction to the Maximum
Integral Flow problem.

\subsection{Reduction to Maximum Integral Flow}

Informally, an integral flow network consists of a directed graph $N$, together with a
single \emph{source vertex} $r$ and a single \emph{sink vertex} $s$\,, and where each arc
$x \arc y$ in $N$ has an associated capacity $c(x \arc y) \in \N$\,. We are interested
in sending come measurable commodity from the source to the sink, under the constraints that
\vspace{0.5ex}
\begin{romanum}
\setlength\itemsep{0.5ex}
\item
	the commodity can only be sent between vertices in the direction of an arc;
\item
	any amount of the commodity which enters a vertex other than the sink
	must also leave that vertex;
\item
	the amount of commodity sent over an arc is at most the capacity of the arc.
\end{romanum}
\vspace{0.5ex}
In particular, we are interested in possible \emph{network-flow functions} $u$\,,
where $u(a) \in \N$ gives the amount of the commodity which we send across the arc $a$\,.
Define
\begin{align}
		F_u(x)
	\;\;=&\;\;
		\sqparen{\sum_{(w \arc x)} u(w \arc x)}
		\;\;-\;\;
		\sqparen{\sum_{(x \arc y)} u(x \arc y)}:
\end{align}
then, the conditions above state that $F_u(x) = 0$ for $x \notin \brace{r,s}$\,, and that for
each arc $a$\,, we have $u(a) \le c(a)$\,. The Maximum Integral Flow problem is to find a
network-flow $u$ such that $F_u(s)$ is maximal: this has several known efficient solutions
which are commonly implemented in programming libraries. We will not discuss them here except
for the purpose of run-time analysis: interested readers may refer to \cite{CLRS} or \cite{CCPS}.

To reduce finding a maximum family of vertex-disjoint $I$--\;\!$O$ paths to Maximum Integral
Flow, we use the following standard network construction:
\DE
	\label{dfn:maxFlowNetwork}
	Let a geometry $(G, I, O)$ be fixed. Define a digraph $N$ with the vertices
	\begin{itemize}
	\setlength\itemsep{0.5ex}
	\item
		$\mA = \set{A_v}{v \in I \union O^c}$ \emph{(out-flow vertices)};
	\item
		$\mB = \set{B_v}{v \in O \union I^c}$ \emph{(in-flow vertices)},
		such that $A_v = B_v$ iff $v \in I \inter O$\,;
	\item
		a \emph{source} vertex $r$ and a \emph{sink} vertex $s$,
	\end{itemize}
	and with the following arcs:
	\begin{itemize}
	\setlength\itemsep{0.5ex}
	\item
		For $i \in I$, $A_i$ has only the incoming arc $r \arc A_i$\,; all other
		in-take vertices $A_v$ have only the incoming arc $B_v \arc A_v$\,.
	\item
		For $\omega \in O$, $B_\omega$ has only the outgoing arc $B_\omega \arc s$\,;
		all other out-take vertices $B_v$ have only the outgoing arc $B_v \arc A_v$\,.
	\item
		If $v \in O^c$ and $w \in I^c$\,, there is an arc $A_v \arc B_w$ iff
		$vw \in G$\,.
	\end{itemize}
	We call $N$ the \emph{max-flow digraph} of $(G,I,O)$\,.
\ED
The purpose of defining this network is to force any network flow function $u$ on the
arcs of $N$ to describe a collection of paths from $r$ to $s$ in $D$\,, from which we
can obtain a family of vertex-disjoint $I$--\;\!$O$ paths in $G$\,. We will now describe
how this construction can be used to find such a maximum family of paths.

For the case where $I$ and $O$ are disjoint, consider the network that would be obtained
by adding source and sink vertices $r$ and $s$ to $V$\,, replacing each edge $xy \in E$ with
arcs $x \arc y$ and $y \arc x$\,, adding arcs $r \arc i$ and $\omega \arc s$ for all $i \in I$
and $\omega \in O$\,. If we set the capacities of all arcs in $N$ equal to $1$\,, then a
maximum integral flow in $N$ could be represented by a set $S$ of arcs $a$ in $N$ for which
$u(a) = 1$\,. From the condition $F_u(x) = 0$ for $x \notin \brace{r,s}$\,, each vertex
other than $r$ or $s$ would have the same number of arcs in $S$ which enter it as it does
that leaves it. Then, the arcs in $S$ describe a collection of \emph{edge-disjoint} paths
from $r$ to $s$\,, and this collection of paths is at a maximum when $u$ is a maximum integral
flow.
 
To force \emph{vertex}-disjointness, we want at most one arc of $S$ to enter each vertex,
and at most one arc of $S$ to leave each vertex. We can do this by replacing each vertex
$x$ by two vertices $B_x \arc A_x$\,, where all arcs which left $x$ now leave $A_x$\,, and
all arcs which entered $x$ now enter $B_x$\,. Because only one arc leaves $B_x$\,, at most
one arc of $S$ can enter $B_x$\,; and similarly with arcs leaving $A_x$\,. Having a maximum
family of vertex-disjoint paths in $N$ from $r$ to $s$\,, we can recover a maximum family of
vertex-disjoint paths in $G$ by removing $r$ and $s$\,, and contracting all pairs $\brace{B_x,
A_x}$ to a single vertex $x \in V$\,. 

For the case where $I$ and $O$ are not disjoint, we can let $A_x$ and $B_x$ be equal
when $x \in I \inter O$\,, as long as we do not connect $x$ to any vertices other
than $r$ or $s$\,. The network that results will be the max-flow digraph $N$ described
in Definition~\ref{dfn:maxFlowNetwork} above.

Given a geometry $(G,I,O)$ the max-flow digraph for $(G,I,O)$ can easily be constructed
in $O(m)$ time. Then, by the above construction, we can attempt to find a path cover for
$(G,I,O)$ as follows:
\AL
	\label{alg:findCover}
	\emph{Input:} the max-flow digraph $N$ of a geometry $(G,I,O)$\,, with
	arc capacities set to $1$ for each arc of $N$\,.
	\\\emph{Output:} either \ri~a~path cover $\mC$ for $(G,I,O)$ and a function
	$f: O^c \to I^c$ whose orbits induce the paths in $\mC$\,, or \rii~no~flow
	exists for $(G,I,O)$\,.
	\begin{enumerate}
	\item
		Set $u$ to be a maximum integral network flow for $N$\,.
	\item
		Find a family $\mC$ of directed paths in $G$ which start at input vertices, end
		at output vertices, and only traverses edges $xy$ such that $u$ is non-zero on
		the arc $A_x \arc B_y$\,.
	\item
		For each arc $A_x \arc B_y$ of $\mC$\,, set $f(x) = y$\,.
		\\---~~\ri~If the paths of $\mC$ cover all vertices of $G$\,, return
				$(\mC, f)$\,.
		\\---~~\rii~Otherwise, no flow exists for $(G,I,O)$.
	\end{enumerate}
\LA

Whether or not $(G,I,O)$ has a flow, the collection of paths $\mC$ which is constructed
will be a maximum family of vertex-disjoint $I$--\;\!$O$ paths; we can check if they
cover $V$ (and construct the function $f$) by simply traversing the paths. By
Theorem~\ref{thm:causalUniquePathFamily}, if $\mC$ doesn't cover all vertices,
then $(G,I,O)$ has no causal path cover, in which case it also has no flow.

Let $n = \abs{V}$\,, $m = \abs{E}$\,, and $k = \abs{I} = \abs{O}$\,. The Maximum Integral
Flow problem on $N$ can be solved in time $O(km)$ by the Ford-Fulkerson algorithm (see,
\eg~\cite{CLRS})\,. The time required to construct $\mC$ and $f$\,, and to verify that
$\mC$ is a path cover, will be $O(n)$\,. Thus the running time of Algorithm~\ref{alg:findCover}
is $O(km)$ when $G$ is a connected graph.

\subsection{Obtaining the Dependency Order of $f$}

The function $f$ returned as part of Algorithm~\ref{alg:findCover} is necessary,
but not sufficient, for $(G,I,O)$ to have a flow. We must also discover if $f$
has a compatible partial order which satisfies \flowii\ and \flowiii. Suppose that
$f$ has such a compatible partial order, and consider in particular the dependency
order $\preceq$ of $f$\,. Because it is the coarsest compatible order, $\preceq$ can
be defined as the transitive closure of the binary relation $\del$ on $V$\,, given by
\begin{align}
		x \del y
	\quad\implies\quad
		[y = x]
		\;\;\vee\;\;
		[y = f(x)]
		\;\;\vee\;\;
		[y \sim f(x)]	\;,
\end{align}
which we will call the \emph{dependency relation}: we will say that $y$ \emph{depends} on $x$
if $x \del y$\,. We can then reduce finding the dependency order to the Transitive Closure
problem, which is concerned precisely with finding the transitive closure of binary
relations.\footnote{The Transitive Closure problem is usually expressed in terms of
directed graphs, but doing so doesn't provide any additional insight into the problem,
and is at any rate equivalent to the problem for binary relations.} If there is a cycle
of dependency relations, then the transitive closure of $\del$ will not be a partial order,
in which case $f$ is not a flow function; otherwise, $f$ is a flow function, and the
transitive closure will be the dependency relation of $f$\,.

Algorithm~\ref{alg:findDepthFn} is a Tarjan-like algorithm for detecting cycles in the
dependency relation $\del$ (given the function $f$ and graph $G$ which determines $\del$),
and computing the transitive closure if no such cycles exist. It also takes the path cover
$\mC$ as input.
\AL
	\label{alg:findDepthFn}
	\emph{Input:} A geometry $(G,I,O)$\,, and the output $(\mC, f)$ of
	Algorithm~\ref{alg:findCover}.
	\\\emph{Output:} either \ri~the~dependency order $\preceq$ of $f$,
	or \rii~no~flow exists for $(G,I,O)$\,.
	\begin{itemize}
	\item
		Let \typ{sup}\ be an array mapping pairs $(x, P) \in V \x \mC$ to the
		supremum (\ie\ the least upper bound) of the set $\brace{x}$ in the path $P$\,,
		or the empty set if there is no least upper bound.\footnote{In the literature
		for the Transitive Closure problem, this is known as a \emph{chain decomposition}
		of the directed graph represented by the relation $\del$\,.} Initialize
		$\typ{sup}[x,P] = \vide$ for all $(x, P) \in V \x \mC$\,.

	\item
		Let \typ{status}\ be an array mapping $x \in V$ to one of
		$\brace{\typ{none}\,,\; \typ{pending}\,,\; \typ{fixed}}$\,;
		where $\typ{status}[x] = \typ{none}$ indicates that no supremum
		of $\brace{x}$ has yet been discovered, and $\typ{status}[x] =
		\typ{fixed}$ indicates that all suprema of $\brace{x}$
		have been discovered. Initialize $\typ{status}[x] = \typ{none}$ for
		all $x \in V$\,.
	
	\item
		For each vertex $x \in V$ with $\typ{status}[x] = \typ{none}$ perform the
		following recursive algorithm to set the suprema of $\brace{x}$\,:
		\begin{enumerate}
		\item
			Set $\typ{status}[x] = \typ{pending}$\,.
		\item
			For the path $P_x$ which contains $x$\,, set $\typ{sup}[x,P_x] = x$\,.
		\item
			For each vertex $y$ which depends on $x$ (aside from $x$ itself):
			\begin{itemize}
			\item
				if $\typ{status}[y] = \typ{none}$\,, recursively determine the
				suprema of $\brace{y}$\,.
			\item
				if after this $\typ{status}[y] \ne \typ{fixed}$\,,
				$(G,I,O)$ has no flow; abort.
			\item
				otherwise, set $\typ{sup}[x,P] = \min(\,\typ{sup}[x,P] \,,\;
				\typ{sup}[y,P]\,)$\; for each path $P \in \mC$ (taking the minimum
				with respect to distance from the start of $P$).
			\end{itemize}
		\item
			Set $\typ{status}[x] = \typ{fixed}$\,.
		\end{enumerate}
	\item
		Once $\typ{status}[x] = \typ{fixed}$ for all $x \in V$\,, return the array
		$\typ{sup}$\,.
	\end{itemize}
\LA
This algorithm determines if there is a cycle of dependencies by performing a depth-first
search, and checking if it traverses a sequence of relations $x \del y \del \cdots \del x$ in
the course of finding the elements which are greater than $x$ in the dependency order. If it
does, it will discover that $\typ{status}[x] = \typ{pending}$ on the second time it visits
$x$\,: then, it may as well abort, as $f$ cannot be a flow function and no dependency order
exists.

The array $\typ{sup}$ returned as output characterizes the dependency order $\preceq$\,:
we have $x \preceq y$ if and only if $\typ{sup}[x,P_y] \le y$\,, where $P_y$ is the path
containing the vertex $y$\,, and the inequality $\le$ is with respect to the distance from
the initial vertex of $P_y$\,. We can compute this easily by constructing functions which
maps each vertex $y \in V$ to an index for $P_y$\,, and another function mapping each vertex
$x \in V$ to the distance of $x$ from the initial point of $P_x$\,. This can be considered
as a part of building the path cover $\mC$ in Algorithm~\ref{alg:findCover} (indeed, such
functions characterize the path cover $\mC$ and may be considered to be how $\mC$ is
represented).

Let $n = \abs{V}$\,, $m = \abs{E}$\,, and $k = \abs{I} = \abs{O} = \abs{\mC}$\,. The time
required to fix the values of $\typ{sup}[x,P]$ for each path $P \in \mC$ is $O(k \deg(x))$\,.
Summing over all $x \in V$\,, the cost of recursively obtaining the suprema for all of $V$ is
$O(km)$\,. This is then the cost of Algorithm~\ref{alg:findDepthFn} as a whole, as initializing
$\typ{sup}$ and $\typ{status}$ takes a total of $O(kn) \subset O(km)$ time for $G$ connected.

\section{Discussion and Conclusion}

We have presented a decomposition of unitary maps into three successive operations $P$, $\Phi$, and $R$. The first one represents the familiar preparation map and expands the computation space by introducing auxiliary qubits; the second one is diagonal in the computational basis and has only unit coefficients; and the last is a restriction map that contracts back the computational space into the chosen output space. It is important to emphasize that both $R$ and $P$ have a very simple structure, and hence the decomposition suggests that the whole quantum computing part of an algorithm is encoded in the phase map operator. 

The restriction map $R$ does not correspond directly to a meaningful physical transformation, so this decomposition is a mathematical artifact. However, we have illustrated how one may attempt to find a one-way patterns which effectively implements such phase decompositions. We have also shown how one may attempt to find a phase map decomposition directly. This constitutes a method for the implementation of unitary maps in the one-way model which completely bypasses the circuit model. It is hoped that the phase map decomposition may produce more efficient implementations in particular instances. 

An interesting restriction which may be imposed is to restrict the choice of the allowed angles. For example, we might restrict to multiples of $\pi/4$\,, in the hope of staying within the
world of Pauli measurements and the so-called $\pi/4$ magic preparations, which may help with implementations \cite{dk05a}. In this restricted case, the Equations of Lemma~\ref{L-eq} may no longer have an exact solution in general; however, one does not expect an exact solution in this case, since the obtained model is only approximately universal. It would be interesting to see whether one can compute down an approximate phase map decomposition such that $U\approx R \circ \Phi \circ P$ with arbitrary precision. 

Another question is whether there is a reasonable relaxation of the definition of a pattern, by allowing new commands which would allow additional possible phase maps.  How much a non-trivial extension would complicate the theory (\eg, standardization, flow) and the implementation (\eg, requiring only one simple primitive for entanglement) of one-way patterns is not clear.

\bibliographystyle{unsrt}
\bibliography{elham_new}

\begin{thebibliography}{10}

\bibitem{Gott99}
D.~Gottesman and I.~L. Chuang.
\newblock Quantum teleportation is a universal computational primitive.
\newblock {\em Nature}, 402:390, 1999.

\bibitem{RB01}
R.~Raussendorf and H.-J. Briegel.
\newblock A one-way quantum computer.
\newblock {\em Physical Review Letters}, 86(5188), 2001.

\bibitem{mqqcs}
R.~Raussendorf, D.~E. Browne, and H.-J. Briegel.
\newblock Measurement-based quantum computation on cluster states.
\newblock {\em Phys. Rev. A}, 68(022312), 2003.

\bibitem{Nielsen03}
M.~A. Nielsen.
\newblock Universal quantum computation using only projective measurement,
  quantum memory, and preparation of the 0 state.
\newblock {\em Phys. Lett. A.}, 308:96, 2003.

\bibitem{Schl03}
D.~Schlingemann.
\newblock Cluster states, algorithms and graphs.
\newblock quant-ph/0305170, 2003.

\bibitem{graphstates}
M.~Hein, J.~Eisert, and H.-J. Briegel.
\newblock Multi-party entanglement in graph states.
\newblock {\em Phys. Rev. A}, 69:62311, 2004.

\bibitem{DAB03}
W.~{D\"{u}r}, H.~Aschauer, and H.-J. Briegel.
\newblock Multiparticle entanglement purification for graph state.
\newblock {\em Phys. Rev. Lett.}, 91:107903, 2003.

\bibitem{VDD04b}
M.~Van den Nest, J.~Dehaene, and B.~De Moor.
\newblock Graphical description of the action of local clifford transformations
  on graph states.
\newblock {\em Phys. Rev. A.}, 69:0223116, 2004.

\bibitem{VDD04}
M.~Van den Nest, J.~Dehaene, and B.~De Moor.
\newblock An efficient algorithm to recognize local clifford equivalence of
  graph states.
\newblock {\em Phys. Lett. A.}, 70:034302, 2004.

\bibitem{MP04}
M.~Mhalla and S.~Perdrix.
\newblock Complexity of graph state preparation.
\newblock quant-ph/0412071, 2004.

\bibitem{L03}
D.~W. Leung.
\newblock Quantum computation by measurements.
\newblock {\em IJQI}, 2(1), 2004.

\bibitem{AL04}
P.~Aliferis and D.~W. Leung.
\newblock Computation by measurements: a unifying picture.
\newblock {\em Phys. Rev. A}, 70:062314, 2004.
\newblock quant-ph/0404082.

\bibitem{CLN04}
A.~M. Childs, D.~W. Leung, and M.~A. Nielsen.
\newblock Unified derivations of measurement-based schemes for quantum
  computation.
\newblock quant-ph/0404132, 2004.

\bibitem{PJ04}
S.~Perdrix and P.~Jorrand.
\newblock Measurement-based quantum turing machines and their universality.
\newblock quant-ph/0404146, 2004.

\bibitem{JP04}
P.~Jorrand and S.~Perdrix.
\newblock Unifying quantum computation with projective measurements only and
  one-way quantum computation.
\newblock quant-ph/0404125, 2004.

\bibitem{DKP04}
V.~Danos, E.~Kashefi, and P.~Panangaden.
\newblock The measurement calculus.
\newblock quant-ph/0412135, 2004.

\bibitem{dk05a}
V.~Danos and E.~Kashefi.
\newblock Pauli measurements are universal.
\newblock In Peter Selinger, editor, {\em Proceedings of the 3rd International
  Workshop on Quantum Programming Languages, QPL 2005, Chicago}. ENTCS, June
  2005.

\bibitem{RV05}
T.~Rudolph and S.~S. Virmani.
\newblock A relational quantum computer using only two-qubit total spin
  measurement and an initial supply of highly mixed single qubit states.
\newblock quant-ph/0503151, 2005.

\bibitem{RAB04}
R.~Raussendorf, Simon Anders, and H.-J. Briegel.
\newblock Fault-tolerant quantum computation using graph states.
\newblock Communication to the Quantum Information and Quantum Control
  Conference, Fields Institute, Toronto.
  http://atlas-conferences.com/c/a/n/n/80.htm, July 2004.

\bibitem{ND04}
M.~A. Nielsen and C.~M. Dawson.
\newblock Fault-tolerant quantum computation with cluster states.
\newblock quant-ph/0405134, 2004.

\bibitem{RBH04}
R.~Raussendorf, S.~Bravyi, and J.~Harrington.
\newblock Long-range quantum entanglement in noisy cluster states.
\newblock {\em Phys. Rev. A.}, 71:062313, 2004.

\bibitem{PL05}
P.~Aliferis and D.~W. Leung.
\newblock Fault-tolerant quantum computation in the graph-state model.
\newblock quant-ph/0503130, 2005.

\bibitem{VBR05}
M.~Varnava, D.~E. Browne, and T.~Rudolph.
\newblock Loss tolerant one-way quantum computation -- a horticultural
  approach.
\newblock quant-ph/0507036, 2005.

\bibitem{RHG05}
R.~Raussendorf, J.~Harrington, and K.~Goyal.
\newblock fault-tolerant one-way quantum computer.
\newblock quant-ph/0510135, 2005.

\bibitem{DHN05}
C.~M. Dawson, H.~L. Haselgrove, and M.~A. Nielsen.
\newblock Noise thresholds for optical quantum computers.
\newblock quant-ph/0509060, 2005.

\bibitem{Nielsen04}
M.~A. Nielsen.
\newblock Optical quantum computation using cluster states.
\newblock {\em Phys. Rev. Lett.}, 93(4):040503, 2004.

\bibitem{CMJ04}
S.R. Clark, C.~Moura Alves, and D.~Jaksch.
\newblock Efficient generation of graph states for quantum computation.
\newblock {\em New J. Phys.}, 7(124), 2005.

\bibitem{BR04}
D.~E. Browne and T.~Rudolph.
\newblock Resource-efficient linear optical quantum computation.
\newblock {\em Phys. Rev. Lett.}, 95:010501, 2005.

\bibitem{TPKV04}
M.~S. Tame, M.~Paternostro, M.~S. Kim, and V.~Vedral.
\newblock Toward a more economical cluster state quantum computation.
\newblock quant-ph/0412156, 2004.

\bibitem{TPKV05}
M.~S. Tame, M.~Paternostro, M.~S. Kim, and V.~Vedral.
\newblock Natural three-qubit interactions in one-way quantum computing.
\newblock quant-ph/0507173, 2005.

\bibitem{nature05}
P.~Walther, K.~J. Resch, T.~Rudolph, E.~Schenck, H.~Weinfurter, V.~Vedral,
  M.~Aspelmeyer, and A.~Zeilinger.
\newblock Experimental one-way quantum computing.
\newblock {\em Nature}, 434:169--176, March 2005.

\bibitem{KPA05}
A.~Kay, J.~K. Pachos, and C.~S. Adams.
\newblock Graph state preparation and cluster computation with global
  addressing of optical lattices.
\newblock quant-ph/0501166, 2005.

\bibitem{BES05}
S.C. Benjamin, J.~Eisert, and T.M. Stace.
\newblock Optical generation of matter qubit graph states.
\newblock {\em New J. Phys.}, 7:194, 2005.

\bibitem{CCWD05}
Q.~Chen, J.~Cheng, K.~Wang, and J.~Du.
\newblock Efficient construction of 2-d cluster states with probabilistic
  quantum gates.
\newblock quant-ph/0507066, 2005.

\bibitem{BBFM05}
S.~C. Benjamin, D.~E. Browne, J.~Fitzsimons, and J.~J.~L. Morton.
\newblock Brokered graph state quantum computing.
\newblock quant-ph/0509209, 2005.

\bibitem{dk05c}
V.~Danos and E.~Kashefi.
\newblock Determinism in the one-way model.
\newblock quantph-0506062, 2005.

\bibitem{generator04}
V.~Danos, E.~Kashefi, and P.~Panangaden.
\newblock Robust and parsimonious realisations of unitaries in the one-way
  model.
\newblock {\em Phys. Rev. A}, 72, 2005.

\bibitem{Beaudrap06}
N.~de~Beaudrap.
\newblock Characterizing {\protect \&} constructing flows in the one-way
  measurement model in terms of disjoint {$\protect I$}--\;\!{$\protect O$}
  paths.
\newblock quant-ph/0603072, 2006.

\bibitem{CLRS}
T.~H. Cormen, C.~E. Leiserson, R.~L. Rivest, and C.~Stein.
\newblock {\em Introduction to Algorithms}.
\newblock MIT Press and McGraw-Hill, 2001.
\newblock 2nd edition.

\bibitem{CCPS}
W.~J. Cook, W.~H. Cunningham, W.~R. Pulleybank, and A.~Schrijver.
\newblock {\em Combinatorial Optimization}.
\newblock Wiley-Interscience New York, 1998.
\newblock Section~3.2, pages~38--45.

\end{thebibliography}

\end{document}